\documentclass[10pt,conference]{IEEEtran}
\usepackage{cite}
\usepackage{amsmath,amssymb,amsfonts}
\usepackage{algorithmic}
\usepackage{graphicx}
\usepackage{textcomp}
\usepackage{xcolor}
\usepackage{physics}
\usepackage{caption}
\usepackage{subcaption}
\usepackage{array}
\usepackage{comment}
\usepackage{url}
\def\BibTeX{{\rm B\kern-.05em{\sc i\kern-.025em b}\kern-.08em
    T\kern-.1667em\lower.7ex\hbox{E}\kern-.125emX}}
\graphicspath{ {./figures/} }
\begin{document}

\title{Asymmetry of CNOT gate operation in superconducting transmon quantum processors using cross-resonance entangling\\
}

\author{\IEEEauthorblockN{Travis Hurant}
\IEEEauthorblockA{\textit{Department of Electrical and Computer Engineering} \\
\textit{IBM Q Hub at NC State}\\
\textit{NC State University}\\
Raleigh, USA \\
tahurant@ncsu.edu}
\and
\IEEEauthorblockN{Daniel D. Stancil}
\IEEEauthorblockA{\textit{Department of Electrical and Computer Engineering} \\
\textit{IBM Q Hub at NC State}\\
\textit{NC State University}\\
Raleigh, USA \\
ddstanci@ncsu.edu}
}

\maketitle

\begin{abstract}
Controlled-NOT (CNOT) gates are commonly included in the standard gate set of quantum processors and provide an important way to entangle qubits. For fixed-frequency qubits using the cross-resonance entangling technique, using the higher-frequency qubit to control the lower-frequency qubit enables much shorter entangling times than using the lower-frequency qubit as the control. Consequently, when implementing a CNOT gate where logical control by the lower-frequency qubit is needed, compilers may implement this functionality by using an equivalent circuit such as placing Hadamard gates on both qubits before and after a CNOT gate controlled by the higher-frequency qubit. However, since the implementation is different depending on which qubit is the control, a natural question arises regarding the relative performance of the implementations. We have explored this using quantum processors on the IBM Q network. The basic circuit used consisted of operations to create a Bell State, followed by the inverse operations so as to return the qubits to their initial state in the absence of errors (Hadamard + CNOT + barrier + CNOT + Hadamard). The circuit depth was varied using multiples of this basic circuit. An asymmetry in the error of the final state was observed that increased with the circuit depth. The strength and direction of the asymmetry was unique but repeatable for each pair of coupled qubits tested. This observation suggests that the asymmetry in CNOT implementation should be characterized for the qubits of interest and incorporated into circuit transpilation to obtain the best accuracy for a particular computation.
\end{abstract}


\section{Introduction}
Experimental quantum computing has made great strides in the past few decades. From the basic quantum logic gates realized by Monroe et al. \cite{monroe:1995} to the 32 quantum volume machines recently announced by IBM \cite{IBM:QV32}, quantum computing systems have grown in size, computational power and general interest. Publications on new or improved algorithms point to the potential of quantum computing. Yet, several key obstacles must be overcome if today’s envisioned applications become tomorrow’s reality. One such obstacle is the infidelity of two-qubit gates.

Two-qubit gate operations are fundamental to quantum computing. It has been shown that a quantum universal gate set can be made with one and two-qubit gates \cite{divincenzo:1995} \cite{barneco:1995} and as a result two-qubit gates are ubiquitous in quantum algorithms. 
However, two-qubit gate fidelity must be sufficiently high to ensure quantum algorithms yield useful results.

%
Characterizing two-qubit gate errors is therefore paramount to obtaining high quality computational results
from NISQ era quantum processors. Gate error data is used to evaluate the general performance and hence selection of a quantum computing system. Furthermore, it is used in compiler optimization techniques such as noise-aware circuit compilation \cite{murali:2019} and variability-aware qubit allocation and movement \cite{tannu:variability:2019}. Inaccuracies in the characterization of gate operation can yield unexpected results and poor computational performance. 

At IBM the Controlled-NOT (CNOT) gate is characterized by a single error rate. However, an exploration of the pulse-level quantum controls used by IBM to execute a CNOT gate uncovered two different microwave pulse schedules for 
a CNOT realized with a given pair of qubits, depending on which qubit is the control. 
If the error characteristics of the CNOT configurations differ, then the systems, controls and compilers that make use of this data may not be fully optimized. 

In this paper, we explore the performance of the different CNOT configurations and attempt to quantify an observed asymmetry. Our results suggest that taking this asymmetry into account when mapping a circuit onto physical gates could reduce the error in the computation.

\section{Background}
IBM uses superconducting transmon quantum processors with fixed frequency qubits, and uses the cross-resonance (CR) gate as a two-qubit entangling gate \cite{chow:2011}.
The CR gate stimulates a $ZX$ interaction by driving the control qubit with microwave pulses at the frequency of the target qubit \cite{alexander:2020}.
To generate a CNOT gate, the CR gate is combined with an additional frame change $S$ on the control qubit and a \(X_{\frac{\pi}{2}}\)on the target qubit \cite{IBM:faq}. IBM combines the CNOT gate with the single qubit $U$ gate, defined by \(U(\theta, \psi, \lambda) = R_z(\phi)R_y(\theta)R_z(\lambda)\), to form a universal quantum gate set \cite{zulehner:2019}. 
\begin{figure}[b]
\begin{subfigure}[t]{0.24\textwidth}
\centering
\includegraphics[width=0.7\textwidth]{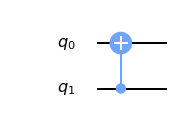}
\caption{}
\label{fig:nonoptimal cnot}
\end{subfigure}
\begin{subfigure}[t]{0.24\textwidth}
\centering
\includegraphics[width=0.7\textwidth]{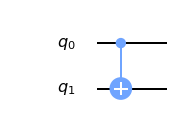}
\caption{}
\label{fig:optimal cnot}
\end{subfigure}
\begin{subfigure}[t]{0.48\textwidth}
\centering
\includegraphics[width=0.4\textwidth]{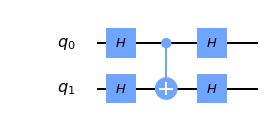}
\caption{}
\label{fig:inverted cnot}
\end{subfigure}
\caption{Three examples of a CNOT gate. The circuits in (a) and (c) are equivalent. If a circuit is programmed to execute the $\mathrm{CNOT_{1,0}}$ seen in
part (a)
but it is not physically allowed, a compiler may swap it out for the circuit in part (c).
This effectively takes the $\mathrm{CNOT_{0,1}}$ to the $\mathrm{CNOT_{1,0}}$; the gate originally intended by the circuit designer.}
\label{fig:cnot replacements}
\end{figure}
%

In general, the CR gate used in the IBM quantum processors is restricted to use the higher frequency qubit as the control for the lower frequency qubit, since this results in a shorter entangling time \cite{tripathi:2019} \cite{magesan:2020} \cite{IBM:backend}. However, this restriction is not enforced at the software level and circuit designers are free to configure the CNOT gate to use the lower frequency qubit as the control. To overcome this conflict, circuit or microwave pulse compilers replace the disallowed CNOT with an equivalent gate or pulse sequence that adheres to the underlying physical restriction. An example of one such gate sequence can be seen in FIG. \ref{fig:inverted cnot}. This means that the gate sequence for a CNOT may vary with the configuration of the control and target qubits. If these varied gate sequences have different error characteristics, it is plausible that CNOT performance may vary with configuration as well.


We endeavored to explore this idea within the IBM quantum system. However, before we could begin, we first needed to verify whether the microwave pulse sequences used in the IBM control systems varied with CNOT configuration. Using Qiskit \cite{Qiskit}, an open source software development kit used for working with the IBM quantum processors, we built two circuits. The first contained a single CNOT gate configured to use the higher frequency qubit as the control. The second circuit contained a single CNOT gate configured to use the lower frequency qubit as the control. We then used the \texttt{schedule()} \cite{IBM:schedule} function available in Qiskit to generate a pulse schedule for each circuit (using the \texttt{ibmq\char`_almaden} quantum processor as the backend). As expected, we found the schedules to differ, as detailed in TABLE \ref{table:cnot pulse comparison}. With this established, we then moved on to designing our experiments.

\begin{table}[t]
\centering
    \begin{tabular}{|m{2cm}|m{2cm}|m{2cm}|}
        \hline
        \textbf{CNOT Configuration} & \textbf{Number of Gates} & \textbf{Schedule Duration}  \\
        \hline
        $\mathrm{CNOT_{hfreq}}$ & 13 & 348 ns \\
        \hline
        $\mathrm{CNOT_{lfreq}}$ & 21 & 384 ns \\ 
        \hline
    \end{tabular}
    \caption{The number of gates and the duration of the pulse schedule was dependent on CNOT configuration. When the CNOT was configured to use the lower frequency qubit as the control ($\mathrm{CNOT_{lfreq}}$) the schedule included more gates and had a longer duration.}
    \label{table:cnot pulse comparison}
\end{table}


%

\section{Methods}   

\begin{figure}[b]
    \centering
    \includegraphics[width=0.40\textwidth]{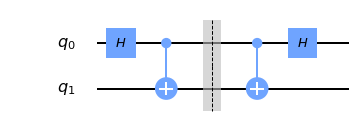}
    \caption{The identity operation is composed of a Bell state and its inverse. The $n$-stage circuit consists of $n$ repetitions of this identity operation. }
    \label{fig:identity operation}
\end{figure}

To evaluate the symmetry of a CNOT, we conducted an experiment which consisted of multiple circuits, with each circuit consisting of $n$ repetitions of an identity operation, where $n= 1,2, …, 6$. The identity operation, which can be seen in FIG. \ref{fig:identity operation}, is composed of a Bell State and its inverse. A barrier is inserted to ensure the circuit is not converted to an identity gate during compilation. 
This experiment was run twice for each pair of qubits included in our evaluation. In the first run 
the CNOT gates were configured to use $q_0$ as the control and $q_1$ as the target. In the second run 
the CNOT gates were configured to use $q_1$ as the control and $q_0$ as the target.
Note, here $q_0$ and $q_1$ are being used generically to represent two different qubits and do not explicitly represent the qubits indexed with 0 or 1 on any particular quantum processor. 

\begin{figure*}[t]
    \centering
    \begin{subfigure}{0.9\textwidth}
        \centering
        \includegraphics[width=0.9\textwidth]{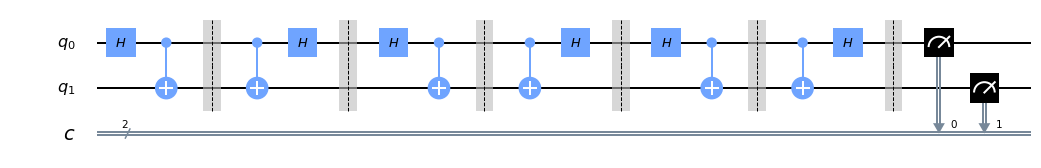}
        \caption{}
        \label{fig:3-stage q0_1}    
    \end{subfigure}
    \begin{subfigure}{0.9\textwidth}
        \centering
        \includegraphics[width=0.9\textwidth]{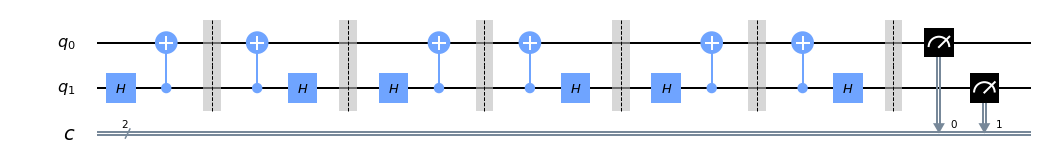}
        \caption{}
        \label{fig:3-stage q1_0}    
    \end{subfigure}
    \caption{In the 3-stage circuit, the identity operation is repeated 3 times. Barriers are placed between the identity operations and between the Bell state and its inverse to prevent compiler optimizations that may alter the gate sequence. The direction of the CNOT and placement of the Hadamard gates are reversed in \ref{fig:3-stage q0_1} and \ref{fig:3-stage q1_0} to evaluate the effect of CNOT direction on circuit performance.}
    \label{fig:3-stage circuit}
\end{figure*}

In this paper we refer to the circuit with $n$ repetitions of the identity operation as the $n$-stage circuit. The 3-stage circuits can be seen in FIG. \ref{fig:3-stage circuit}. Note that barriers are inserted between each stage to ensure the compiler does not combine identity operations.

For each circuit in the experiment, the selected qubits were first prepared in the \(\ket{00}\) state after which the $n$ identity operations of the $n$-stage circuit were run. In the ideal scenario each identity operation would return the qubits to the \(\ket{00}\) state, and therefore we would also expect to measure the qubits in the \(\ket{00}\) state at the end of the $n$-stage circuit. However, given the presence of noise and systematic errors in the physical system, actual results varied and included all states in the computational space. We benchmarked the performance of each CNOT direction on a qubit pair by calculating the percentage of shots that were measured in the \(\ket{00}\) state. A CNOT that exhibited symmetry in implementation direction should produce ground state percentages invariant of implementation direction. Even in the presence of noise and systematic error we expect the percentages to be statistically similar if the CNOT is symmetric. 

We define the asymmetry of an $n$-stage circuit as the absolute magnitude of the difference between the fraction of correct answers for each implementation:

\begin{equation}\label{asymmetry}
    f(n) = \left|g_{01}(n) - g_{10}(n)\right|,
\end{equation}
\begin{equation}\label{percent in ground}
    g_{c,t}(n) = \frac{G_{c,t}(n)}{T_{c,t}(n)},
\end{equation}
where \textit{\(G_{c,t}(n)\)} and \(T_{c,t}(n)\) are the number of shots measured in the ground state and the total number of shots, respectively, for the $n$-stage circuit with CNOTs configured with \textit{c} as the control and \textit{t} as the target.

We classified the CNOT pair as asymmetric if $f(n)\geq 0.02$ for any value of $n$ considered (i.e., for $n$ between 1 and 6). Since measurement errors are typically of order 2-3\%, this is roughly equivalent to classifying the CNOT pair as asymmetric if the asymmetry is larger than the read error.

Specific qubit pairs on each processor were evaluated using the experiment described above. The two qubits used in each experiment were nearest neighbors and were connected by a bus resonator to ensure SWAP gates were not included during circuit compilation. Each $n$-stage circuit was executed 3 times with a shot count of 4,096 shots. Results were aggregated so the total shot count for an $n$-stage circuit was 12,288.

Experiments were conducted on \texttt{ibmq\char`_valencia}, \texttt{ibmq\char`_almaden} and \texttt{ibmq\char`_paris}, which are 5, 20 and 27 qubit processors, respectively. Our work focused primarily on \texttt{ibmq\char`_almaden} on which we tested every qubit pair in the coupling map. To determine if the asymmetry was present on other quantum processors we ran additional experiments on \texttt{ibmq\char`_valencia} and \texttt{ibmq\char`_paris}. We were able to confirm the presence of CNOT asymmetry on these machines through a limited number of experiments on a subset of the available qubit pairs.
The qubit pairs evaluated on each machine can be found in Appendix \ref{appendix:experiments}.

\section{Results}

\subsection{\texttt{ibmq\char`_almaden}}
 An asymmetry was observed in 71.8\% of the experiments conducted on \texttt{ibmq\char`_almaden}. Of the 23 qubit couplings available on \texttt{ibmq\char`_almaden}, 21 exhibited an asymmetry at some point in our evaluation. In 9.6\% of tests, the asymmetry was seen in the shallow depth 1-stage circuit.  The 6-stage circuit, the deepest circuit tested, exhibited the asymmetry in 67.12\% of tests. The divergence in performance was most pronounced for the 6-stage circuits which on average showed a 5.94\% difference in the fraction of correct answers for the different CNOT implementations. 
 Table \ref{tab:almaden results} displays the percentage of experiments in which an asymmetry was observed for each $n$-stage circuit.


\begin{table*}[t]
    \centering
    \begin{tabular}{|c|c|}
    \hline
        \textbf{Circuit} & \textbf{Percentage of Experiments Exhibiting an Asymmetry} \\
         \hline
        1-stage circuit & 9.6\% \\
        \hline
        2-stage circuit & 30.14\% \\
        \hline
        3-stage circuit & 38.36\% \\
        \hline
        4-stage circuit & 52.05\% \\
        \hline
        5-stage circuit & 63.01\% \\
        \hline
        6-stage circuit & 67.12\% \\
        \hline
    \end{tabular}
    \caption{The percentage of experiments that exhibited an asymmetry for each of the $n$-stage circuits. The asymmetry increased with circuit depth.} 
    \label{tab:almaden results}
\end{table*}

\begin{figure}[h]
    \centering
    \begin{subfigure}{0.49\textwidth}
        \centering
        \includegraphics[width=0.8\textwidth]{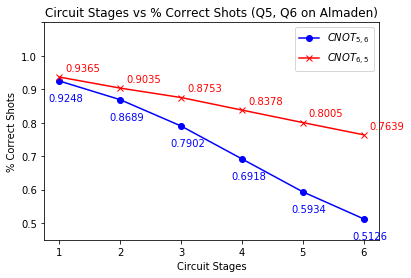}
        \caption{}
        \label{fig:almaden 5 and 6}    
    \end{subfigure}
    \begin{subfigure}{0.49\textwidth}
        \centering
        \includegraphics[width=0.8\textwidth]{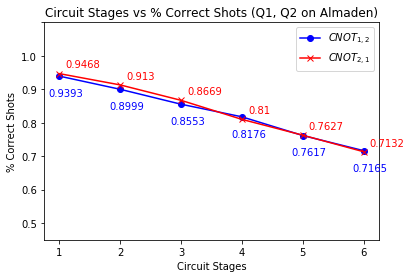}
        \caption{}
        \label{fig:almaden 1 and 2}    
    \end{subfigure}
    \caption{An asymmetry of the CNOT gate was observed across multiple pairs of qubits on \texttt{ibmq\char`_almaden}. The asymmetry increased with circuit depth as seen in (a). The performance of CNOT gates which did not exhibit an asymmetry was invariant to CNOT configuration as seen in (b). }
    \label{fig:cnot asymmetry}
\end{figure}

In Fig. \ref{fig:cnot asymmetry} we plot the fraction of correct answers for each CNOT implementation as a function of circuit stage. Fig. \ref{fig:almaden 5 and 6} illustrates the asymmetric performance of a CNOT on qubits 5 and 6 on \texttt{ibmq\char`_almaden}. The fraction of correct answers diverges substantially as $n$ increases. The asymmetric behavior can even be seen at the shallow-depth 2-stage circuit. This suggests the possibility of non-optimal performance for circuits which incorporated the \(\text{CNOT}_{5,6}\) gate on \texttt{ibmq\char`_almaden} on the day this experiment was conducted.

In contrast, Fig. \ref{fig:almaden 1 and 2} shows the performance of a CNOT gate configured to use qubits 1 and 2 on \texttt{ibmq\char`_almaden} that did not display an asymmetry. In this plot we do not see a divergence in the fraction of correct answers and performance of the \(\text{CNOT}_{1,2}\) on \texttt{ibmq\char`_almaden} on the day of experimentation appeared to be invariant to CNOT configuration. 

Interestingly, the CNOT asymmetry was even observed on qubits with relatively low error rates. In one such example we tested qubits 3 and 4 on \texttt{ibmq\char`_almaden}. Daily calibration data published by IBM on the day of experimentation showed the error rates for qubits 3 and 4 to be some of the lowest on the \texttt{ibmq\char`_almaden} quantum processor (see Appendix \ref{appendix:calibration data} for specific values). Again, the percentage of shots measured in the ground state diverged as $n$ increased, an indicator of the performance asymmetry. 
This suggests that CNOT asymmetry characterization would be a useful addition to the calibration data.

\begin{figure}[h]
    \centering
    \includegraphics[width=0.45\textwidth]{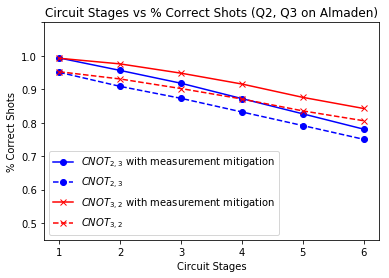}
    \caption{Measurement error mitigation improved overall CNOT performance, but did not reduce CNOT asymmetry.}
    \label{fig:measurement error mitigation}
\end{figure}

A performance difference was also observable when measurement error mitigation techniques were applied. On IBM quantum processors, read operations are typically more error prone than single-qubit and two-qubit gates \cite{tannu:2019}. To determine if read errors factored into the observed asymmetry, we applied a measurement error mitigation technique and compared the mitigated results to the raw results. For these tests we used the technique described in \cite{IBM:measurement}. In general, measurement error mitigation did not adequately address CNOT asymmetry, and in some cases even exacerbated the disparity in performance between CNOT configurations. Fig. \ref{fig:measurement error mitigation} illustrates one example. In this case, measurement error mitigation improved overall CNOT performance which can be seen in the vertical translation of the mitigated plot relative to the unmitigated plot. However, the asymmetry increased when the measurement error mitigation technique was applied. Specifically, the absolute magnitude defined in Eq. \ref{asymmetry} increased from 5.55\% to 6.18\%, constituting an 11.35\% increase in asymmetry. 

\subsection{\texttt{ibmq\char`_valencia} and \texttt{ibmq\char`_paris}}

\begin{figure}
    \centering
    \begin{subfigure}{0.49\textwidth}
        \centering
        \includegraphics[width=0.7\textwidth]{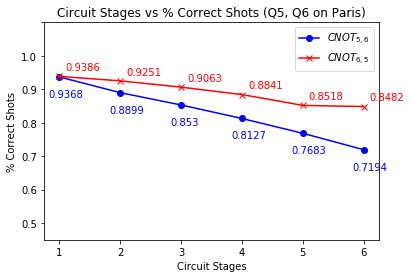}
        \caption{}
        \label{fig:paris 5 and 6}    
    \end{subfigure}
    \begin{subfigure}{0.49\textwidth}
        \centering
        \includegraphics[width=0.7\textwidth]{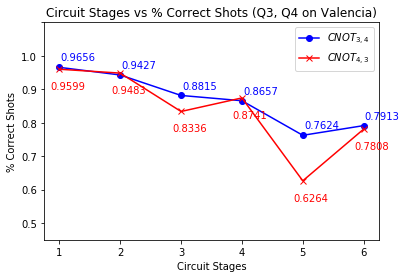}
        \caption{}
        \label{fig:valencia 3 and 4}    
    \end{subfigure}
    \caption{An asymmetry of the CNOT gate was also observed across multiple pairs of qubits on \texttt{ibmq\char`_valencia} and \texttt{ibmq\char`_paris}. The asymmetry generally increased with circuit depth, although CNOTs using Q3 and Q4 on \texttt{ibmq\char`_valencia} exhibited a non-monotonic dependence on circuit depth, as shown in part (b).}
    \label{fig:my_label}
\end{figure}

The asymmetry was also observable on \texttt{ibmq\char`_valencia} and \texttt{ibmq\char`_paris}, despite a smaller sample size
when compared to \texttt{ibmq\char`_almaden}. We observed an asymmetry on several CNOT gates, one of which can be seen in  Fig. \ref{fig:paris 5 and 6}.  The CNOT gate configured to use qubits 5 and 6 on \texttt{ibmq\char`_paris} displayed an asymmetry for $n\geq2$.

The pattern of asymmetry observed on \texttt{ibmq\char`_valencia} was markedly different than the asymmetry observed on other machines. A periodicity can be seen in Fig. \ref{fig:valencia 3 and 4} which is not present on any of the other plots. This pattern was observed only on the Q3, Q4 pair
and suggests the asymmetry may be due to a systematic error.


\section{Conclusions and Future Work}

In this work we presented experimental evidence showing an asymmetry in CNOT operation. Specifically, the performance of the CNOT gate on the IBM Q processors may vary depending upon the configuration of the control and target qubits. Some qubit pairs showed what appears to be a periodic variation in the asymmetry, suggesting the presence of systematic rather than purely random errors.

Based on these findings, we suggest two separate CNOT error characterizations, one for each configuration of control and target qubits, be considered when evaluating CNOT gate operation.  This additional information would better characterize the true performance of a CNOT gate and provide insightful statistics for research into quantum controls and compiler optimizations. 

Our results suggest that
noise-aware circuit compilation can be improved by incorporating configuration-specific CNOT error rates into optimization objective functions. Prior work in noise-aware quantum circuit compilation has incorporated daily calibration data into optimization techniques for qubit mapping \cite{murali:2019}. The efficacy of these techniques is dependent on the accuracy of the reported gate errors. Since current calibration data from IBM characterizes CNOT gate errors with a single error rate, these prior works assume the CNOT gate operation to be invariant to control and target configuration. Our work shows this is not always the case. 
Therefore we believe if stated objective functions are updated to include multiple CNOT error rates, the performance of noise-aware compilers can be further improved. 

Possibilities for future work include additional analysis of the CNOT gate and the underlying CR gate to better understand the underpinnings of the observed asymmetry. In particular, performing state tomography on the output of the multi-stage circuit may give insight into the source of both systematic and random errors. 
Access to the IBM Q Network processors was obtained through the IBM Q Hub at NC State.

\bibliographystyle{./bibliography/IEEEtran}
\bibliography{./bibliography/IEEEabrv,./bibliography/main}
\clearpage
\begin{appendices}



\newpage

\section{Experiments}
\label{appendix:experiments}

\begin{table}[h]
\centering
    \begin{tabular}{|m{2cm}|m{2cm}|m{2cm}|m{2cm}|}
        \hline
        &
        \textbf{\texttt{ibmq\char`_almaden}} &\textbf{\texttt{ibmq\char`_valencia}}& \textbf{\texttt{ibmq\char`_paris}} \\
        \hline
        Total Jobs & 148 & 14 & 26 \\
        \hline
        \multicolumn{4}{|c|}{Tests per Qubit Pair}  \\ 
        \hline
        0,1 & 6 & 1 & 1 \\
        \hline
        1,2 & 4 & 1 & 2 \\ 
         \hline 
        1,4 & -- & -- & 1 \\ 
         \hline 
        1,6 & 5 & -- & -- \\ 
         \hline 
        2,3 & 4 & -- & 1 \\ 
         \hline 
        3,4 & 4 & 5 & -- \\ 
         \hline 
        3,5 & -- & -- & 2 \\ 
         \hline 
        3,8 & 7 & -- & -- \\ 
         \hline 
        4,7 & -- & -- & 1 \\ 
         \hline 
        5,6 & 4 & -- & -- \\ 
         \hline 
        5,8 & -- & -- & 1 \\ 
         \hline 
        5,10 & 6 & -- & -- \\ 
         \hline 
        6,7 & 4 & -- & 1 \\ 
         \hline 
        7,8 & 4 & -- & -- \\ 
         \hline 
        7,10 & -- & -- & 1 \\ 
         \hline 
        7,12 & 2 & -- & -- \\ 
         \hline 
        8,9 & 2 & -- & 2 \\ 
         \hline 
        9,14 & 2 & -- & -- \\ 
         \hline 
        10,11 & 2 & -- & -- \\ 
         \hline 
        11,12 & 2 & -- & -- \\ 
         \hline 
        11,16 & 2 & -- & -- \\ 
         \hline 
        12,13 & 2 & -- & 0 \\ 
         \hline 
        13,14 & 2 & -- & 0 \\ 
         \hline 
        13,18 & 2 & -- & -- \\ 
         \hline 
        15,16 & 2 & -- & -- \\ 
         \hline 
        16,17 & 2 & -- & -- \\ 
         \hline 
        17,18 & 2 & -- & 0 \\ 
         \hline 
        18,19 & 2 & -- & -- \\
        \hline
        Total Pairs Tested & 74 & 7 & 13 \\
        \hline
    \end{tabular}
    \caption{A dash is used to denote qubit-qubit mappings that are not supported for a machine.}
    \label{table:machine tests}
\end{table}

\section{Calibration Data}
\label{appendix:calibration data}
\begin{table}[h]
\centering
    \begin{tabular}{|m{2cm}|m{2cm}|m{2cm}|m{2cm}|m{2cm}|m{2cm}|m{2cm}|}
        \hline
        \multicolumn{7}{|c|}{Error Rates for \texttt{ibmq\char`_almaden}} \\
        \hline
        \textbf{Qubit} & U1 & U2 & U3 & $\mathrm{CNOT_{3,4}}$ & $\mathrm{CNOT_{4,3}}$ & Read  \\
        \hline
        Q3 & 0.0 & 0.00042 & 0.00084 & 0.00862 & -- & 0.025 \\
        \hline
        Q4 & 0.0 & 0.00041 & 0.00082 & -- & 0.00862 & 0.035 \\ 
        \hline
    \end{tabular}
    \caption{Daily gate error rate data for qubits 3 and 4 on \texttt{ibmq\char`_almaden}. On the date published, these error rates were some of the lowest reported for \texttt{ibmq\char`_almaden}.}
    \label{table:almaden 2 and 3 error rates}
\end{table}

\end{appendices}
\end{document}